\documentclass[final,3p,times]{elsarticle}
\usepackage{hyperref}

\usepackage{amssymb}

\usepackage{tabulary}

\journal{Current Opinion in Structural Biology}

\begin{document}

\begin{frontmatter}



\title{Adaptive machine learning for protein engineering}


\author[inst1,inst2]{Brian L. Hie}

\affiliation[inst1]{organization={Department of Biochemistry, Stanford University School of Medicine},
            city={Stanford},
            state={CA},
            postcode={94305}, 
            country={USA}}
\affiliation[inst2]{organization={Stanford ChEM-H, Stanford University},
            city={Stanford},
            state={CA},
            postcode={94305}, 
            country={USA}}

\author[inst3,cor1]{Kevin K. Yang}

\ead{yang.kevin@microsoft.com}
\fntext[cor1]{Corresponding author.}
\affiliation[inst3]{organization={Microsoft Research New England},
            city={Cambridge},
            state={MA},
            postcode={02142}, 
            country={USA}}

\begin{abstract}
Machine-learning models that learn from data to predict how protein sequence encodes function are emerging as a useful protein engineering tool.
However, when using these models to suggest new protein designs, one must deal with the vast combinatorial complexity of protein sequences.
Here, we review how to use a sequence-to-function machine-learning surrogate model to select sequences for experimental measurement.
First, we discuss how to select sequences through a single round of machine-learning optimization.
Then, we discuss sequential optimization, where the goal is to discover optimized sequences and improve the model across multiple rounds of training, optimization, and experimental measurement.
\end{abstract}




\end{frontmatter}


\section{Introduction}
\label{sec:intro}
 
Protein engineering seeks to design or discover novel proteins with useful properties~\cite{Arnold2018}, but doing so is challenging because (i) evaluating a given design is costly (requiring a laboratory experiment to express and characterize each design) and (ii) the search space of all possible protein sequences is very large (with twenty natural amino acids, there are $20^{100}$ possible sequences of length 100, a search space larger than the number of atoms in the universe)~\cite{Yang2019}.

To reduce the number of evaluations required, researchers can leverage machine learning to train a surrogate model that predicts the property of interest from the protein sequence.
Because making predictions from a machine-learning model is less expensive and faster than conducting a wetlab experiment, the surrogate model can reduce the overall experimental burden~\cite{Yang2019, Fox2007}.
However, once a sequence-to-function model is available, a second consideration remains: how does one select new designs from the combinatorially large protein search space?

In this review, we address this question through what we call ``adaptive machine learning,'' which covers two problem settings.
First, we consider the problem of using a trained machine-learning surrogate model to select one or more optimized sequences for laboratory measurement.
Second, we consider the problem of finding optimized sequences over sequential rounds of experimental measurement, model training, and property prediction.
We describe key principles, highlight useful examples from the literature, and discuss areas that would benefit from future methodological improvements.

\section{Overview and key principles}

\begin{figure}
    \centering
    \includegraphics[width=\textwidth]{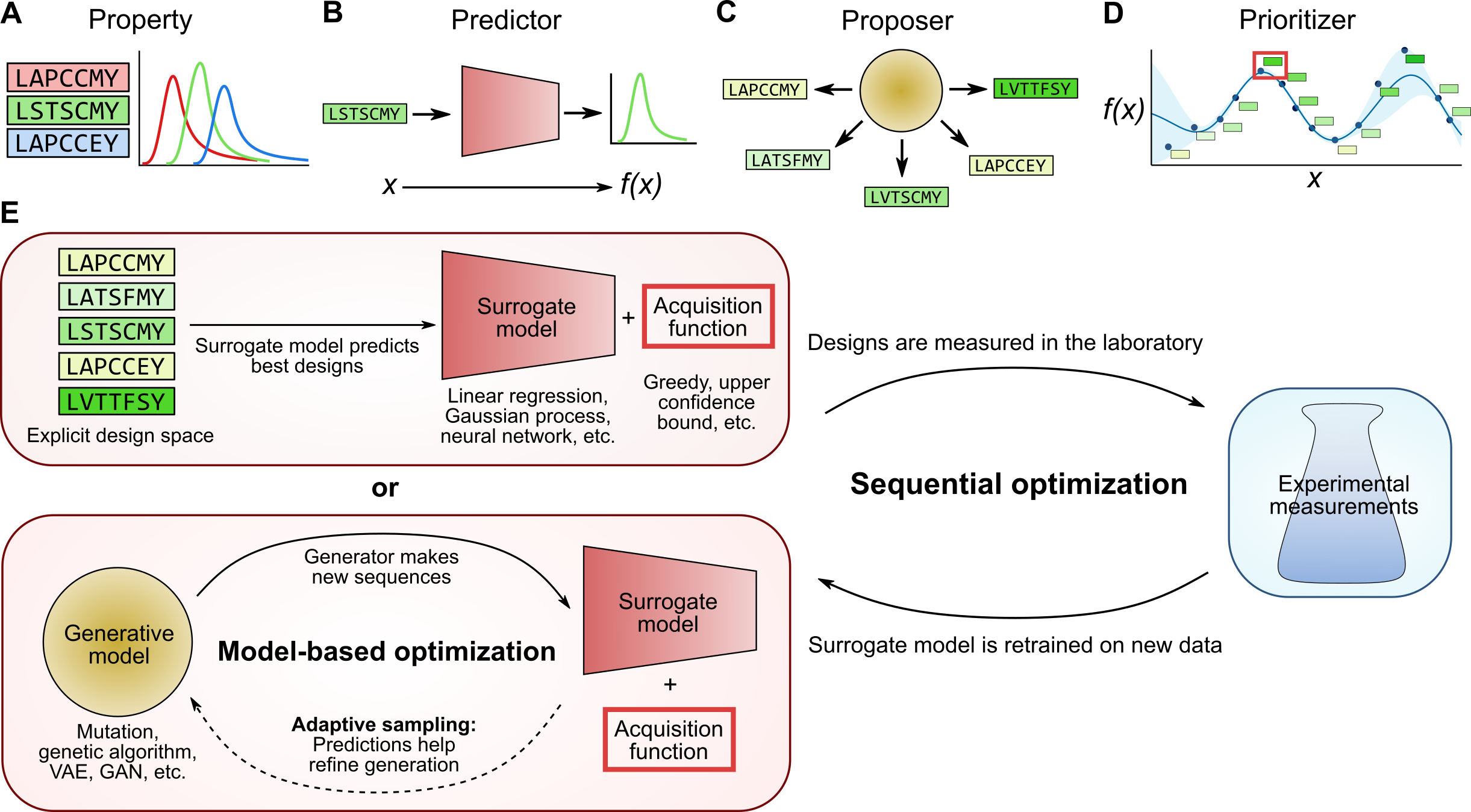}
    \caption{Overview of adaptive machine learning for protein engineering. Four key components are the optimization property, the surrogate model that predicts the property given a sequence, a generative model that proposes sequences, and an acquisition function that prioritizes sets of sequences to measure given sequence information and predictions from the surrogate model. Sequences can be acquired from an explicit design space or from sequences proposed by a generative model. Optionally, the surrogate model's predictions can guide the generative model toward better designs through adaptive sampling. Acquired sequences can be measured in the laboratory and the resulting data can help guide future rounds of optimization.}
    \label{fig:concept}
\end{figure}

There are four important components to consider when doing adaptive learning for protein optimization: the property to optimize, the sequence-to-function predictor, the acquisition function that prioritizes designs, and the generative model that proposes the designs (Figure~\ref{fig:concept}A-D).

\subsection{The property}

The \textbf{optimization property} is the phenotype-of-interest (Figure~\ref{fig:concept}A).
For example, one could maximize the fluorescence intensity at a given excitation wavelength.
Multiple properties may be considered, such as maximizing the enzymatic activity while minimizing the immunogenicity of a protein drug.

\subsection{The predictor}
The \textbf{surrogate model} takes in protein sequence information and predicts the value of the optimization property (Figure~\ref{fig:concept}B).
The surrogate model requires a number of design considerations, including how to represent the protein sequence and what kind of machine-learning algorithm to use.
Typical sequence representations range from a simple binary encoding of the raw sequence to more complex continuous neural encodings~\cite{Wittmann2021,Frappier2021}. 
Model architectures also range in complexity from linear regression to deep neural networks, and have been reviewed in-depth previously~\cite{Yang2019,Frappier2021}.
Often, it is also desirable for the surrogate model to quantify the uncertainty in its predictions in order to make reliable decisions. 
Popular models that provide a notion of uncertainty include Gaussian processes and model ensembles, which we review in Section~\ref{sec:outer}.

\subsection{The prioritizer}

\begin{figure}
    \centering
    \includegraphics[width=\textwidth]{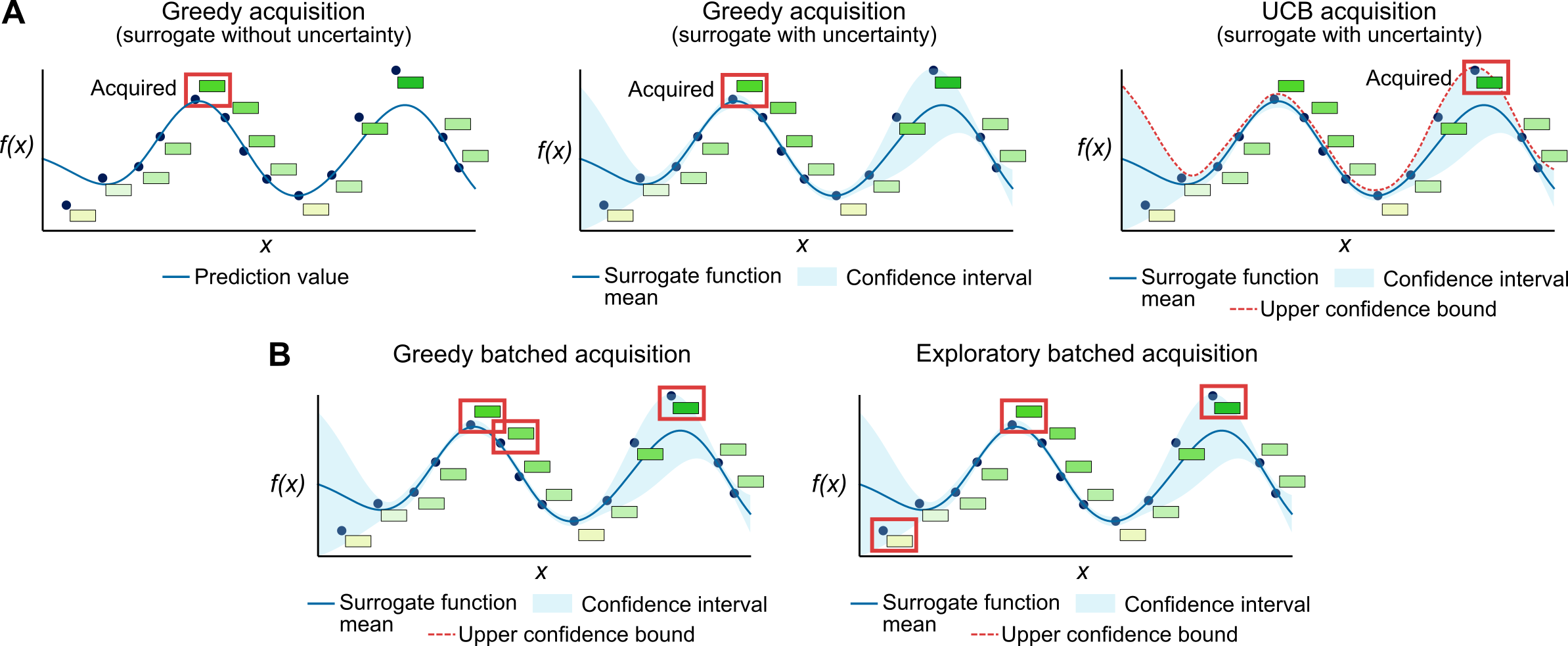}
    \caption{Overview of acquisition functions. (\textbf{A}) Greedy acquisition can work well but may limit search to suboptimal regions of the sequence landscape, whereas incorporating uncertainty helps an acquisition function explore the landscape. (\textbf{B}) Greedy batched acquisition can acquire similar examples, whereas more diverse batches provide a better picture of the global sequence landscape.}
    \label{fig:acquisition}
\end{figure}

An \textbf{acquisition function} uses sequence information and the output of the surrogate model to prioritize designs for experimental measurements (Figure~\ref{fig:concept}D).
The simplest acquisition function greedily selects the top prediction (or the top few predictions) according to the surrogate model.
Greedy acquisition is common in practice and can work well~\cite{Fox2007,Wu2019,Wittmann2020,Biswas2021,Bryant2021,Singer2021}, but can also limit search to narrower regions of the protein landscape (Figure~\ref{fig:acquisition}A), which we discuss further in Section~\ref{sec:outer}.

Many standard acquisition functions are designed to select only a single example on each round.
Often, however, obtaining many experimental measurements in parallel is desirable.
While simply acquiring several of the top-ranked sequences is possible, this approach may be prone to acquiring many similar examples (Figure~\ref{fig:acquisition}B).
Special methods for batched acquisition are therefore designed to encourage acquiring more diverse sequences~\cite{Azimi2010,Romero2013,Gonzalez2016,Yang2020,Sinai2020}, but this remains an open area of methodological development.

\subsection{The proposer}

Although evaluating the surrogate model is faster and cheaper than actually obtaining laboratory measurements, it is nevertheless impossible to evaluate the surrogate on all possible protein sequences.
One approach is to explicitly define a \textbf{design space} of proteins.
Example design spaces include traditional library designs for protein engineering such as all single- or double-mutants, all the possible mutations at a small number of sites~\cite{Wu2019,Wittmann2020,Wu2016}, or a recombination library defined by mixing and matching parts of homologous parent sequences~\cite{Romero2013,Voigt2002,Otey2006,Smith2013}.

Another approach is to use a \textbf{generative model} to implicitly define the design space by learning a probability distribution over sequences (Figure~\ref{fig:concept}C).
A generative model performs one or both of two fundamental tasks: (i) assigning every possible sequence a likelihood of being in the proposal distribution, and (ii) generating examples of sequences from a proposal distribution.
The simplest models assume this distribution can be modeled by considering sites independently or relationships between pairs of sites~\cite{Hopf2017EVmutation,Russ2020}.
More recently, researchers have used \textbf{deep neural network generative models} to learn more complex sequence distributions and propose sequences for evaluation~\cite{Wu2021}; relevant neural architectures include variational autoencoders (VAEs)~\cite{Kingma2014,Riesselman2018,Brookes2018}, generative adversarial networks (GANs)~\cite{Goodfellow2020,Gupta2019}, and autoregressive language models~\cite{Hochreiter1997,bepler2019ssa,Shin2021,Madani2021}.

\begin{table}
\begin{tabulary}
    {\linewidth}{|C|C|C|C|C|C|C|}
    \hline
        \textbf{Reference and date} & \textbf{Optimization property} & \textbf{Surrogate model} & \textbf{Acq. func.} & \textbf{Generative model / design space} & \textbf{Seq. opt.?} & \textbf{In-vitro?} \\ \hline
        Fox et al.\ (2007)~\cite{Fox2007} & Enzyme catalytic activity & Linear regression & Greedy & Sequence recombination & Yes & Yes \\ \hline
        Romero et al.\ (2013)~\cite{Romero2013} & Protein thermostability & Gaussian process & UCB & Sequence recombination & Yes & Yes \\ \hline
        Bedbrook et al.\ (2017)~\cite{Bedbrook2017} & Protein localization & Gaussian process & UCB & Sequence recombination & Yes & Yes \\ \hline
        Wu et al.\ (2019)~\cite{Wu2019} & Enzyme catalytic activity & Regressor ensemble & Greedy & Explicitly-defined design space & Yes & Yes \\ \hline
        Brookes et al.\ (2019)~\cite{Brookes2019} & Protein fluorescence & Neural network ensemble & Greedy & Neural network (VAE) & No & No \\ \hline
        Kumar and Levine (2019)~\cite{Kumar2019} & Protein fluorescence & Neural network ensemble & Greedy & Neural network (GAN) & No & No \\ \hline
        Gupta and Zou (2020)~\cite{Gupta2019} & Antimicrobial activity & Neural network (RNN) & Greedy & Neural network (GAN) & No & No \\ \hline
        Liu et al.\ (2020)~\cite{Liu2020} & Antibody affinity & Neural network ensemble & Greedy & Activation maximization & No & Yes \\ \hline
        Wittmann et al.\ (2020)~\cite{Wittmann2020} & Protein expression and binding & Regressor ensemble & Greedy & Explicitly-defined design space & Yes & No \\ \hline
        Anishchenko et al.\ (2020)~\cite{Anishchenko2020} & Valid folding & Neural network (CNN) & Greedy & Sequence mutation & No & Yes \\ \hline
        Biswas et al.\ (2021)~\cite{Biswas2021} & Protein fitness, fluorescence & Linear regression & Greedy & Sequence mutation & No & Yes \\ \hline
        Bryant et al.\ (2021)~\cite{Bryant2021} & Protein viability & Classifier ensemble & Greedy & Sequence mutation & No & Yes \\ \hline
        Greenhalgh et al.\ (2021)~\cite{Greenhalgh2021} & Enzyme catalytic activity & Gaussian process & UCB & Sequence recombination & Yes & Yes \\ \hline
\end{tabulary}
\caption{Examples of adaptive learning for protein engineering. Acq.\ func.: Acquisition function. Seq.\ opt.: Sequential optimization, indicates studies that performed multiple rounds of variant selection and surrogate model training. In-vitro: Indicates studies that obtained in-vitro measurements of new protein sequences. RNN: Recurrent neural network~\cite{Hochreiter1997}. CNN: Convolutional neural network~\cite{Fukushima1980}.}
\label{tab:examples}
\end{table}

\section{Model-based protein optimization}

After obtaining a surrogate model, \textbf{model-based optimization} can prioritize sequences of interest by finding the sequence or sequences in the design space that optimize the acquisition function (Figure~\ref{fig:concept}E).
Notably, in this setting we do not provide the surrogate model with new training data.
The simplest approach is to first obtain a sequence design space (explicitly-defined or from a generative model) and then to select sequences from this design space based on a surrogate model and a corresponding acquisition function (for example, picking the best predicted sequences in the design space)~\cite{Wu2019,Romero2013,Gomez-Bombarelli2018,Singer2021}.

\subsection{Adaptive sampling from a generative model}
\label{sec:generative}

Rather than separate generation and evaluation steps, the surrogate model can also help shift the generative model so that it proposes more optimal sequences.
For example, in \textbf{adaptive sampling}, protein sequences are sampled from a generative model, the outputs of a surrogate model are used to re-estimate the parameters of the generative model, and the process iterates until convergence~\cite{Brookes2018}.
For example, Brookes and Listgarten use adaptive sampling based on a VAE generative model and a neural network surrogate model to optimize DNA sequences for protein expression abundance~\cite{Brookes2018}.
Gupta and Zou use adaptive sampling based on a GAN generative model and a recurrent neural network surrogate model to design antimicrobial peptides~\cite{Gupta2019}.
To perform \textit{de novo} protein design \cite{Huang2016}, Anishchenko et al.\ use adaptive sampling based on a mutation-based generative model and a neural network surrogate model meant to identify sequences with valid folds~\cite{Anishchenko2020} (Table \ref{tab:examples}).

By default, adaptive sampling assumes a trustworthy surrogate model; however, in practice, these models are imperfect and can be prone to poor predictions in many regions of the protein space.
To address this problem, adaptive sampling can avoid degeneracies in the surrogate model by constraining sampling to be close to the training distribution for the surrogate model~\cite{Brookes2019,Anishchenko2020}.
In each iteration, it is also possible to retrain the surrogate model to avoid pathologies~\cite{Fannjiang2020}.
Sequence generation can also be improved by sampling from an ensemble of generative models~\cite{Angermueller2020P3BO}.

A closely related approach to model-based optimization uses \textbf{genetic algorithms} to heuristically balance both mutation and recombination to produce new sequences~\cite{Hansen2006} by adaptively querying a surrogate model to preserve sequence designs~\cite{Sinai2020}.
Another approach to model-based optimization inverts the surrogate model by finding the elements from the generative model's distribution that are most likely to have a desirable value according to the surrogate model.
An inverse of the surrogate model could be trained via an iterative procedure similar to adaptive sampling~\cite{Kumar2019}, or the inverse of a differentiable surrogate model can be computed using gradient-based methods~\cite{Liu2020,Linder2020a,Linder2020b}.

\section{Sequential optimization}
\label{sec:outer}

In sequential optimization, the surrogate model has access to multiple rounds of experimental measurement, which provide new data that can be re-incorporated into subsequent rounds of model training~\cite{eisenstein2020active} (Figure~\ref{fig:concept}E).
To train the initial surrogate model, the first batch can consist of random samples from the design space~\cite{Wittmann2020} or sequences with known measurements (for example, from publicly available data)~\cite{Biswas2021}.
Sequential optimization is guided by an \textbf{objective function} that specifies the overall goal of the optimization procedure.
The objective most relevant to protein engineering is to find the protein sequence that maximizes the optimization property (for example, find the most fluorescent sequence).

Greedy acquisition across experimental rounds is the simplest implementation of sequential optimization and is used widely in practice~\cite{Wittmann2020} (Figure~\ref{fig:acquisition}A).
Going beyond greedy acquisition means tolerating more risk for potentially higher reward, which is often described as a tradeoff between ``exploitation'' (equivalent to greedy acquisition) and ``exploration''~\cite{robbins1952some,Auer2003UCB,Snoek2012,sutton2018reinforcement}, in which an algorithm acquires the proteins with high surrogate-model uncertainty to improve future predictions.
Here we focus predominately on Bayesian optimization as the main alternative to greedy acquisition.
Different objectives, such as training the best overall model through active learning~\cite{eisenstein2020active} or reformulating the objective as a reinforcement learning problem~\cite{Angermuller2020} are also possible, though are less common in protein engineering applications.

\subsection{Bayesian optimization}

\textbf{Bayesian optimization} searches for a protein that maximizes the optimization property \cite{Snoek2012} by leveraging Bayesian uncertainty in the surrogate function to guide the exploration-exploitation tradeoff.
Bayesian optimization relies on the acquisition function to systematically weigh the surrogate model's prediction with its associated uncertainty.
The \textbf{upper confidence bound (UCB)} acquisition function adds the prediction value with a weighted uncertainty term that lets the user control the influence of uncertainty on the prediction, with a larger weight encouraging more exploration~\cite{Auer2003UCB,Srinivas2010} (Figure~\ref{fig:acquisition}A).
UCB has good theoretical properties~\cite{Snoek2012} and is used widely in practice~\cite{Romero2013,Bedbrook2017,Greenhalgh2021,hie2020uncertainty} (Table~\ref{tab:examples}).
Other notable acquisition functions select an example that is predicted to, on expectation, have the largest improvement compared to the best example in the training set or compared to a randomly drawn example from the training set~\cite{Snoek2012,Frisby2020,Wilson2018}.
Uncertainty can also help improve exploration in batched acquisition~\cite{Azimi2010,Romero2013,Gonzalez2016,Yang2020} (Figure~\ref{fig:acquisition}B).

\subsubsection{Gaussian-process surrogate models}

A \textbf{Gaussian process} is a popular surrogate model in Bayesian optimization~\cite{Rasmussen2005}.
Gaussian process regression assumes that the optimization-property values of any set of sequences are distributed according to a multivariate Gaussian.
Often, the mean of the distribution is used as the prediction value and the marginal variance can be used to compute uncertainty.
Probabilistic classification with Gaussian processes is also possible but requires approximation~\cite{Kuss2006, Rasmussen2005} and multitask Gaussian processes can be used to predict multiple optimization properties~\cite{Bonilla2009}.
Because of their theoretical elegance, flexibility, and good performance in practice, Gaussian processes have seen wide use in protein engineering applications \cite{Romero2013,Bedbrook2017,Greenhalgh2021,hie2020uncertainty,Bedbrook2019,Voutilainen2020}.

Gaussian processes are defined by a mean function and a kernel function that together specify how training examples influence the prediction.
The most widely used kernel functions~\cite{Micchelli2006} are defined on continuous inputs and can accommodate neural embeddings or binary sequence embeddings~\cite{Wittmann2021,Bedbrook2017,hie2020uncertainty}. 
Kernels can also be defined on discrete inputs, including sequences~\cite{Oh2019,Beck2017,Moss2020}.

The cost of exact inference with Gaussian processes grows cubically with the size of the training data, which may be prohibitive on extremely large protein sequence datasets.
There is a wide literature on scaling Gaussian processes, including methods for exploiting sparsity in the structure of the training data or performing approximate inference, which applies to both continuous and discrete inputs~\cite{liu2020gpscale,Fortuin2019}.

\subsubsection{Beyond Gaussian process regression}

Bayesian optimization can also leverage more bespoke Bayesian models and algorithms for exact or approximate inference~\cite{Koller2009}.
The wide interest in neural network models over the last decade has also led to increased interest in uncertainty prediction through Bayesian neural networks, in which the parameters of the network are themselves random variables with associated prior distributions, though efficient and accurate inference in these models can be challenging~\cite{Neal2012}.

Probabilistic surrogate models can also be implemented by model \textbf{ensembles}, which train multiple sequence-to-function models on the same data and rely on variance in model predictions, due to different model architectures or randomness in the training procedure, to estimate uncertainty~\cite{Liu2020,laks2017ensemble,Zeng2019}.
Ensembles are not Bayesian by default, so incorporating prior information into these models can be challenging~\cite{Amini2019,Izmailov2021}.

\section{Discussion}

Protein engineering is a challenging task given its proven computational hardness~\cite{Pierce2003} and general biological complexity.
Here we have reviewed how machine learning can help protein engineers deal with the immense complexity of the protein sequence landscape by proposing sequence designs and guiding a researcher across multiple rounds of laboratory experimentation.

While the techniques we review are a good representation of the current state-of-the-art, there is much room for methodological development.
When performing Bayesian optimization, obtaining well-calibrated uncertainty estimates can be more difficult when the inputs are discrete or high-dimensional (or both).

Protein engineering also typically involves much fewer rounds (since experimental measurements are often resource-intensive) and much larger batches (for example, using multiplexed experimental designs) than is typically assumed in the theoretical literature for Bayesian optimization.
Another consideration, particularly in scientific applications, is to design proteins without using a natural starting point or for properties that do not exist in natural proteins.
Doing so with data-driven approaches may require additional modeling considerations beyond sequence data alone, such as biophysics, biochemistry, and protein structure.

\section*{Acknowledgements}
We thank Nicholas Bhattacharya and Sam Sinai. for helpful comments and discussion. B.L.H. acknowledges the support of the Stanford Science Fellows program. 


\bibliographystyle{elsarticle-num-annotate} 
\bibliography{cas-refs}





\end{document}